\documentclass[twocolumn,tighten]{aastex62}

\hypersetup{linkcolor=red,citecolor=blue,filecolor=cyan,urlcolor=magenta}

\usepackage{graphicx}
\usepackage{natbib}

\newcommand{\eg}{{\rm e.g.,}}

\newcommand{\ie}{{\rm i.e.,}}

\newcommand{\kms}{\ensuremath{\ \mathrm{km\, s}^{-1}}}

\newcommand{\K}{\, \ensuremath{\rm K}}
\newcommand{\gyr}{\, \ensuremath{\rm Gyr}}
\newcommand{\myr}{\, \ensuremath{\rm Myr}}
\newcommand{\kpc}{\, \ensuremath{\rm kpc}}

\newcommand{\rdisk}{\ensuremath{R_{\mathrm{disk}}}}
\newcommand{\msol}{\ensuremath{\ \mathrm{M_{\odot}}}}

\newcommand{\tform}{\ensuremath{t_\mathrm{form}}}
\newcommand{\timelb}{\ensuremath{t_\mathrm{lookback}}}

\newcommand{\rform}{\ensuremath{R_\mathrm{birth}}}
\newcommand{\fgas}{\ensuremath{f_\mathrm{gas}}}

\newcommand{\varz}{\ensuremath{\sigma^2_z}}
\newcommand{\sigz}{\ensuremath{\sigma_z}}

\newcommand{\vzdisp}{\ensuremath{\sigma_{\mathrm{z}}}}
\newcommand{\vbirth}{\ensuremath{\sigma_{\mathrm{birth}}}}

\newcommand{\vcirc}{\ensuremath{v_{\mathrm{circ}}}}

\newcommand{\sfrden}{\ensuremath{\dot{\Sigma}_{\star}}}

\newcommand{\SIM}{{\tt h277}}

\def\Mo{{\rm M_\odot}}

\newcommand{\gaia}{{\em Gaia}}

\graphicspath{{./}{figures/}}

\shorttitle{The Stellar Age-Velocity Relation}
\shortauthors{Bird et al.}

\begin{document}

\title{Inside Out and Upside-Down:\\ The Roles of Gas Cooling and Dynamical Heating in Shaping the Stellar Age-Velocity Relation}

\correspondingauthor{Jonathan Bird}
\email{jonathan.bird@vanderbilt.edu}

\author{Jonathan C.~Bird}
\affiliation{Department of Physics and Astronomy, Vanderbilt University, 6301 Stevenson Center, Nashville, TN, 37235}

\author{Sarah R.~Loebman}
\altaffiliation{Hubble fellow}
\affiliation{Department of Physics, University of California, Davis, 1 Shields Ave, Davis, CA 95616, USA}

\author{David H.~Weinberg}
\affiliation{Department of Astronomy, The Ohio State University, 140 West 18th Avenue, Columbus, OH 43210}
\affiliation{Center for Cosmology and Astro-Particle Physics, The Ohio State University, 191 West Woodruff Avenue, Columbus, OH 43210}

\author{Alyson Brooks}
\altaffiliation{This simulation is publicly available.  Contact Alyson Brooks \\ for information. abrooks@physics.rutgers.edu}
\affiliation{Department of Physics \& Astronomy, Rutgers University, New Brunswick, NJ, USA}

\author{Thomas R.~Quinn}
\affiliation{Department of Astronomy, University of  Washington, Box 351580, Seattle, WA, 98115}

\author{Charlotte R. Christensen}
\affiliation{Physics Department, Grinnell College, Grinnell, IA, USA}

\begin{abstract}
Kinematic studies of disk galaxies, using individual stars in the Milky Way or statistical studies of global disk kinematics over time, provide insight into how disks form and evolve. We use a high-resolution, cosmological zoom-simulation of a Milky Way-mass disk galaxy (\SIM) to tie together local disk kinematics and the evolution of the disk over time. The present-day stellar age-velocity relationship (AVR) of \SIM\ is nearly identical to that of the analogous solar-neighborhood measurement in the Milky Way. A crucial element of this success is the simulation's dynamically cold multi-phase ISM, which allows young stars to form with a low velocity dispersion (\vbirth$\sim 6 - 8 \kms$) at late times. Older stars are born kinematically hotter (i.e., the disk settles over time in an ``upside-down'' formation scenario), and are subsequently heated after birth.  
The disk also grows ``inside-out'', and many of the  older stars in the solar neighborhood at $z=0$ are present because of radial mixing. We demonstrate that the evolution of \vbirth\  in \SIM\ can be explained by the same model used to describe the general decrease in velocity dispersion observed in disk galaxies from $z\sim 2-3$ to the present-day, in which the disk evolves in quasi-stable equilibrium  and the ISM velocity dispersion decreases over time due to a decreasing gas fraction. 
Thus, our results tie together local observations of the Milky Way's AVR with observed kinematics of high $z$ disk galaxies. \\
\\

\end{abstract}


\section{Introduction}
\label{sec:intro}

With the advent of advanced spectroscopic and IFU surveys, the detailed kinematics and evolution of disk galaxies are now being studied in unprecedented detail \citep[e.g., MaNGA, SAMI, KMOS$^{\rm 3D}$;][]{manga, sami, Wisnioski2019}.
Locally, we can use present-day stellar kinematics to infer the past evolution of our Milky Way (MW) through Galactic Archeology, \citep[e.g.,][]{Belokurov2018,Frankel2018,Frankel2019}.
At higher redshift, large galaxy surveys are yielding results on the evolution of populations of galaxies over time \citep[e.g., see the review by][]{Glazebrook2013}.
If the MW (and other local galaxies) kinematically evolve in a fashion similar to high $z$ galaxies, then in principle results from these two epochs should be able to be explained within the same theoretical framework.  

Galaxies in the Local Group show a clear correlation between stellar age and velocity dispersion \citep{Leaman2017}, e.g., the MW \citep{Casagrande2011}, M31 \citep{Dorman2015}, and M33  \citep{Beasley2015}. 
Observations in all of these galaxies show that the velocity dispersion, $\sigma$, increases with increasing stellar age. Locally, young stars are born dynamically cold \citep[e.g.,][]{Stark1989, Kuhn2019}. 
The classic explanation for this age-velocity relation (AVR) is that it is an evolutionary sequence: stars have always formed in dynamically cold gas, but various gravitational scattering mechanisms steadily increase the random orbital energy as a stellar population ages.  The scattering sources might be Giant Molecular Clouds \citep[GMCs; e.g.,][]{Spitzer1951, Wielen1977, Lacey1984}, or spiral arms and bars \citep[e.g.,][]{Carlberg1985, Sellwood2002, Minchev2006, Roskar2008, Loebman2011, Grand2016}.
An alternative explanation for the origin of the AVR is that older stars were born dynamically hotter in the past, but that the disk ``settles'' and becomes progressively cooler as the star forming gas cools with time.  The stars might be dynamically hotter in the past due to a more active galaxy merger phase \citep{Toth1992, Quinn1993, Brook2004, Martig2014, Hu2018, Buck2020} or due to the fact that gravitational turbulence is higher if the disk is more gas rich \citep{Bournaud2009}, or due to a higher star formation rate \citep[SFR; e.g.,][]{Lehnert2014}. 

The MW presents the best opportunity to infer a detailed understanding of galactic formation.  
Combining proper motions and distances from the \gaia\ survey with stellar properties from spectroscopic surveys yields full 6-D kinematics for hundreds of thousands of stars in the MW's disk along with moderately precise ages \citep{Sanders2018}. Studies of the MW using these data suggest it is possible to recreate the observed AVR with dynamical heating of stars alone to ages of $\sim 8 \gyr$ \citep{Gustafsson2016, Yu2018, Ting2019, Mackereth2019}.
However, these results do not rule out any contribution from disk settling over that same time-span.  In fact, it is clear that steady orbital heating cannot be solely responsible for the relatively high velocity dispersion of stars older than $8 \gyr$ \citep[\eg][]{Lacey1984, Sellwood2014, Mackereth2019}.

Observations of rotating disk galaxies out to $z\sim3$ show increasing velocity dispersion with increasing redshift, \citep[\eg][]{Flores2006, Gnerucci2011, Kassin2012, Wisnioski2015, Simons2016, Mieda2016, Mason2017}.  In these studies, the typical tracer of $\sigma$ is gas that has been heated by star formation \citep[e.g., H$\alpha$ or ionized oxygen;][]{Epinat2009, Forster-Schreiber2009, Lemoine2010}. Notably, molecular and atomic gas show roughly the same evolution, albeit for fewer galaxies \citep[\eg][]{Tacconi2013, Ubler2019}.
Many of the kinematic measurements at high $z$ are for galaxies more massive than the MW (especially at $z>1$), yet disk settling has been found even for less-massive populations \citep{Law2009, Kassin2012, Contini2016, Hirtenstein2019}.  

A number of simulators have examined the origin of the AVR and/or the formation of galaxy thin/thick disks.  One trend that has become clear is that the results can be strongly dependent on the star formation parameters, which are commonly limited by the simulation resolution \citep{House2011, Martig2014, Kumamoto2017}.  By definition, higher resolution simulations can capture higher density gas.  As resolution increases, simulators can restrict star formation to occur in higher density gas that more closely resembles GMCs \citep[e.g.,][]{Governato2010, Dutton2019}.  \citet{House2011} showed that the threshold density for star formation was the dominant parameter in setting the stellar velocity dispersion at birth, \vbirth.  For simulations with a low star formation density threshold ($n \sim 0.1$ cm$^{-3}$) that are unable to resolve cold, dense gas, a dispersion floor is imposed on \vbirth, causing a number of simulations to have stars that maintain a steady \vbirth\ over 8-10 Gyr, and with little to no subsequent heating of the stellar populations after birth.  However, with an increase in resolution and star formation threshold ($n \sim 1-5$ cm$^{-3}$), a different result has emerged: an ``Upside-Down'' formation in which stars form with progressively lower \vbirth\ with time \citep{Bird2013, Grand2016}.  Scattering processes can subsequently heat the orbits of stars, causing the older stars to contribute to the AVR in the solar neighborhood \citep{Brook2012, Bird2013, Aumer2016, Grand2016, Ma2017}. 

The resulting AVRs of some simulated galaxies are very similar in slope and normalization to the MW AVR.  However, the kinematics of the youngest stars are generally too hot relative to observations \citep[\eg][]{House2011, Roskar2013, Bird2013, Martig2014, Grand2016}, even when star formation thresholds with $n > 100$ cm$^{-3}$ are achieved \citep{Buck2020, Sanderson2020}.  In this paper we present the first cosmological simulation of a MW-mass disk galaxy that also reproduces the low \vbirth\ determined for the MW.

Meanwhile, high resolution simulations have separately also been used to interpret the observed decline in $\sigma$ in disk galaxies with time. In general, a higher \vbirth\ in the past is well-motivated on theoretical grounds.  Several groups have found correlations between velocity dispersion and star formation fueled by marginal stability of the disk \citep[\eg][]{BournaudElmegreen2009, Krumholz2012, Forbes2014, Benincasa2016, Krumholz2016, Nestingen2017, Orr2018, Orr2019}. In this framework, gravitational instabilities drive turbulent motions, which then naturally regulate galaxies to be marginally stable. This model predicts a relation between the gas velocity dispersion and the surface density of the gas in the galaxy.  When combined with the Schmidt-Kennicutt star formation relation, the gas surface density is also tied to the star formation rate, \citep[e.g.,][]{Green2014, Leroy2016, Semenov2016}.  If the disk is marginally stable, and it is assumed that gas (rather than stars) is the predominant component of the gravitational potential, the higher gas surface densities and gas fractions at higher redshift \citep[\eg][]{Genzel2006, Daddi2010, Jones2010, Genzel2011, Tacconi2013, Glazebrook2013, Popping2015, Krumholz2016, Stott2016, Wiklind2019} should naturally lead to higher gas velocity dispersion.  For more local galaxies, feedback from star-formation may also play a critical role in setting a dynamical ``floor'' for the star-forming gas \citep[\eg][]{Agertz2009, Faucher-Giguere2013, Stilp2013, Semenov2018, Orr2019, Orr2019b}. These models predict that \vbirth\ should be strongly correlated with the SFR. This correlation has been seen for the population of massive, rotating galaxies observed at high $z$ \citep[\eg][]{Swinbank2012, Wisnioski2015, Turner2017, Ubler2019}. 

While observational works infer disk settling through population studies over redshift, simulators have examined the evolution of individual disks to trace the evolution in $\sigma$ over time \citep{Kassin2014, Ceverino2017, Hung2019}.  In all cases, cosmological zoom simulations of individual disk galaxies show a decrease in gas velocity dispersions since $z \sim1.5$, consistent with observations. \citet{Kassin2014} used four MW-mass simulations (one of which is used in this paper) to demonstrate that individual simulated disks show the same slope of gas velocity dispersion vs $z$ as observations, but they did not investigate the origin of the trend. Likewise, \citet{Hung2019} attempted to measure gas velocity dispersions in a way more directly comparable to IFU measurements in a suite of cosmological zoom disk simulations.  While they find that changes in the dispersion are temporally correlated with both gas accretion onto the galaxy disks and the global SFR, they were unable to draw conclusions about the cause of this correlation. On the other hand, \citet{Ceverino2017} examined the driver of $\sigma$ with $z$. By running a controlled experiment in which feedback was shut off for a short time, \citet{Ceverino2017} found that gravitational instabilities in a marginally stable disk set a floor to the turbulence scale, but that feedback drives gas dispersions higher than gravitational instabilities alone.  Simulations without feedback did not show the declining velocity dispersion with time.  They suggest that the decline is tied to the decreasing gas fractions, and hence SFR, of the simulated disks with time.

In this paper, we use a high-resolution cosmological zoom-in simulation to simultaneously investigate the detailed kinematics of the present-day galaxy as well as the galaxy's kinematic evolution over cosmological time scales.  In fact, this simulation, \SIM,\footnote{We note that a version of \SIM\ was also studied by \citet{House2011}.  However, the version in this paper has twice the spatial resolution, 8$\times$ the mass resolution, and includes an updated ISM and star formation prescription as described in \citet{Christensen2012}.} is one of the simulations studied by \citet{Kassin2014}, and thus it has already been shown that this specific galaxy matches the observed trends of decreasing gas velocity dispersion with time.  We focus on this particular galaxy as it has many properties similar to the MW, including a relatively quiescent merger history and structural parameters within 10\% of the MW at $z=0$ \citep{Loebman2014}. The simulation tracks molecular hydrogen abundances and cooling processes (Section~\ref{sec:sim}) and thus resolves a multi-phase ISM that allows the low \vbirth\ of young stars to be captured.
In Section~\ref{sec:results} we establish that the AVR of \SIM\ at $z=0$ is nearly a perfect match to the MW (Section~\ref{sec:AVR}).  We investigate how \vbirth\ varies with radius and time in Sections \ref{sec:32} and \ref{sec:33}, and show that the same equilibrium model that matches the settling of disk kinematics with $z$ also matches the evolution of \vbirth\ in this individual galaxy (Section~\ref{sec:highzmodel}).  In Section \ref{sec:discussion}, we first discuss the physical processes causing \SIM\ to match the MW's AVR (Section \ref{sec:how}), and then highlight the numerical aspects of the simulation that have allowed these processes to emerge (Section \ref{sec:why}).  We conclude in Section \ref{sec:conclusions}.

\section{The Simulation}
\label{sec:sim}

In this work, we use a cosmologically  derived \citep[WMAP3,][]{Spergel2003}  Milky Way--mass galaxy evolved for $13.7$  Gyr using the parallel $N$--body$+$SPH code {\sc Gasoline} \citep{Wadsley2004}.
The simulation uses a redshift dependent cosmic UV background \citep{Haardt2001} and realistic cooling and heating, including cooling from metal lines \citep{Shen2010} and H$_2$ \citep{Christensen2012}.
Feedback from supernovae type II (SN\,II) is modeled using the ``blastwave'' approach \citep{Stinson2006} in which 10$^{51}$ erg of thermal energy is deposited per SN, and cooling is temporarily disabled based on the local gas characteristics.  SN\,Ia thermal energy is also injected, but cooling is not disabled.
The probability of star formation is a function of the non-equilibrium H$_2$ abundances \citep{Christensen2012}.
The result of tying the star formation to the molecular hydrogen abundance is a greater concentration of the stellar feedback energy and the more efficient generation of outflows.
These outflows ensure that the final galaxy has an appropriate rotation curve \citep{Governato2012,Christensen2014a}, stellar mass fraction \citep{Munshi2013}, metallicity \citep{Christensen2016}, and HI properties \citep{Brooks2017}.

The galaxy we use, \SIM, is a part of the  \textit{g14}  suite of simulations \citep{Christensen2012}, and has been studied extensively elsewhere in the literature \citep[\eg][]{Loebman2012, Loebman2014}, including a demonstration that it has a realistic dwarf satellite population \citep{Zolotov2012, Brooks2014}.  
This galaxy has a gravitational force resolution of 173 pc and mass resolutions of $1.3\times10^5$ $\mbox{$\rm M_{\odot}$}$, $2.7\times10^4$ $\mbox{$\rm M_{\odot}$}$, and $8.0\times10^3$ $\mbox{$\rm M_{\odot}$}$ for the dark matter, gas, and stars, respectively.  It also includes the large-scale environment by using the ``zoom-in'' volume renormalization technique \citep{Katz1993} when creating the initial conditions. 

The virialized halo at every step was identified using Amiga's Halo Finder \citep[AHF;][]{gill04,knollmann09}. AHF calculates $M_{vir}$ as the total mass within a sphere that encloses an overdensity \citep[following][]{Bryan1998} relative to the critical density $\rho_{crit}(z)$.
At redshift zero, \SIM\ has a virial radius $R_{vir}$ of $227$ kpc and $M_{vir}$ of $6.8\times10^{11}$ $\Mo$. Of this, 7\% is in gas, 6\% is in stars, and 87\% is in dark matter. 
A total of $4.6\times10^{6}$ dark matter, $2.1\times10^{6}$ gas and $7.4\times10^{6}$ star particles are  within the virial radius at redshift zero.  

The galaxy has a relatively quiescent merger history, with the last major merger (3:1 halo mass ratio) occuring at $z\sim3$.
At the present day, \SIM\ has an $R$-band bulge to disk ratio of 0.33 \citep[using {\sc Sunrise},][]{Jonsson2006}, and maximum circular velocity of $\sim$235 km/s.   
These structural parameters are within 10\% of those measured for the Milky Way.

\section{Results: The Present-day AVR}
\label{sec:results}

\begin{figure}[tbp]
  {\centering \includegraphics[width=\linewidth]{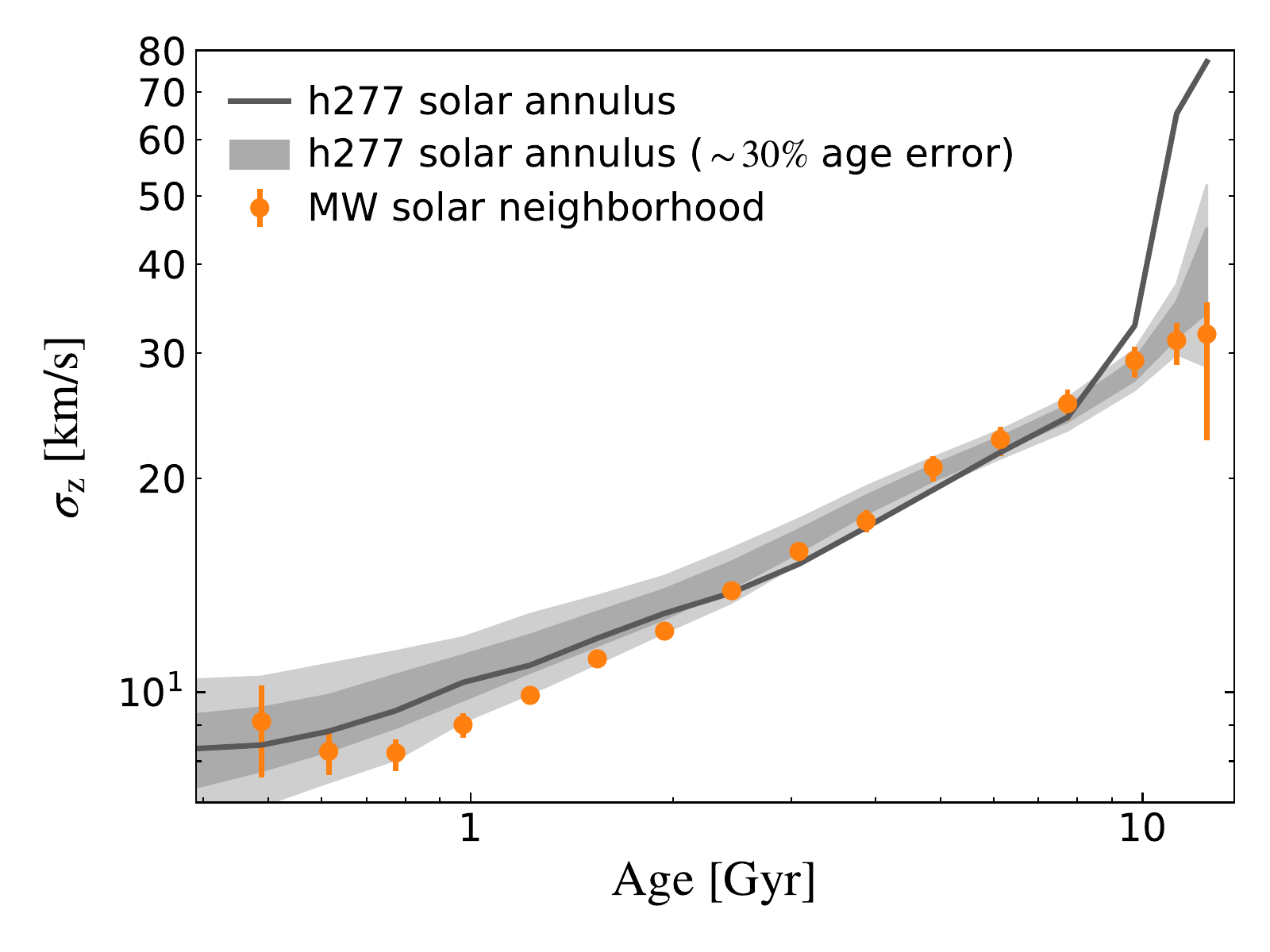}}
  \caption{\label{fig:avr} The vertical velocity dispersion as a function of stellar age in the ``solar neighborhood'' of \SIM\ (black line) compared to the analogous AVR measurement in the Milky Way \citep[orange points, data from][]{Casagrande2011}. The dark and light gray bands represents the $16-84\%$ and $5-95\%$, respectively, range of AVRs that would be found in \SIM\ after accounting for the sample size, age bias, and measurement uncertainty of the re-calibrated GCS survey of the local solar neighborhood. See text for details.}
\end{figure}
 
To analyze the present-day AVR of \SIM, we wish to define a region that is solar neighborhood-like.  \citet{Boardman2020} showed that the scatter in radial gradients of MW analogues is minimized if galaxies are normalized by disk scale length, $R_d$. The best-fit exponential profile to the disk-dominated ($R > 2.5 \kpc$) stellar surface mass density  yields $R_d=2.42$  \kpc.  For the Milky Way, \citet{Bovy13} find $R_d=2.15 \kpc$ for the disk mass surface density profile. The ratio of these scale lengths is equivalent to the ratio of the central radius of the simulated solar annulus to the MW Galacto-centric solar radius, $R_{\odot}$.  Adopting $R_{\odot} = 8 \kpc$, the simulated solar radius is $R = 9 \kpc$ and a solar neighborhood-like annulus spans $8.0 < R/\kpc < 10.0$. We calculate all velocity dispersions using the robust biweight midvariance estimator as implemented in {\tt astropy}\footnote{$\sigma(v) = \sqrt{\zeta_{\mathrm{biweight}}(v)}$. When drawing from a normal distribution with known parameters, we found the biweight midvariance to  outperform the straightforward moment calculation at small sample size ($N < 20$); the two methods are nearly indistinguishable for larger $N$ \citep[see][]{Beers1990}}. 

The present-day age-resolved stellar kinematics of \SIM\ reveal a clear AVR in the disk ($|z| < 1 \kpc$) at the scaled solar annulus (Figure~\ref{fig:avr}). The youngest stars are dynamically cold, with a vertical velocity dispersion \vzdisp\ $\sim 8$ \kms,  which is consistent with the cloud to cloud velocity dispersion measured in molecular gas of GMCs in the MW and nearby galactic disks \citep{Stark1989, Wilson2011} and noteworthy for
a cosmological zoom-in simulation (see Section~\ref{sec:discussion} for discussion).  The velocity dispersion increases with age $\tau$ for older stars; \vzdisp\ gradually increases to $\sim30 \kms$ at $\tau =10 \gyr$ and then quickly rises to $\sim 80 \kms$ for the oldest stars (black line). 

While \SIM\ was not designed to reproduce the MW, the detailed measurements of the MW AVR offer an informative context to interpret our findings.
The intrinsic AVR of \SIM\ (black line in Figure~\ref{fig:avr}) is very similar to the MW AVR, but the oldest stars ($\tau > 9 \gyr)$ are significantly hotter than what is observed.
However, direct comparison of empirical results with the simulated AVR, where all relevant data are known perfectly and completely, can be misleading.
The MW data are from the local solar neighborhood, volume-complete Geneva Copenhagen Survey \citep[GCS;][]{Nordstrom2004, Holmberg2009}. 
We show the MW AVR (orange points) using $N_{\mathrm{GCS}}=3168$ stars of the {\it irfm} sample with ``good ages''  of \citet{Casagrande2011}, the latest re-calibration of the GCS.
Even these high-fidelity data are not perfect: they have a known bias towards younger stars \citep[\eg][]{Nordstrom2004, Casagrande2011}, a mean relative age uncertainty of $28\%$, and are a discrete sampling of the underlying age and kinematic distribution (as $N_{\mathrm{GCS}} \ll$ the number of stars within the solar annulus). 
The simulation data must be ``degraded'' with these sources of noise and bias in order to perform a more legitimate comparison of the \SIM\ and MW AVRs \citep[see also][]{Martig2014, Aumer2016}.

The AVR of \SIM\ is consistent with the MW AVR at all ages after accounting for the measurement noise and bias of the empirical data (grey shaded regions, Figure \ref{fig:avr}).  The \vzdisp\ measured from mock surveys of the \SIM\ solar annulus peak at just $25$ to $40 \kms$ for the oldest stars while maintaining $< 10 \kms$ \vzdisp\ at $\tau < 1 \gyr$ (gray bands in Figure~\ref{fig:avr}). 
Each mock survey contains $N_{\mathrm{GCS}}$ stars randomly drawn from the solar annulus midplane ($8 < R < 10 \kpc$ and $|z| < 1 \kpc$); we then add $0.12$ dex uncertainty in $\log{\tau}$ ($\sim25\%$ relative uncertainty at $\tau=4 \gyr$) and generate the mock ages. Differences in the AVR measured from the intrinsic (black line) and generated mock data (gray regions) are driven by the intrinsic age distribution and the synthetic noise. The intrinsic AVR is usually not at the median of the median observed AVR. The intrinsic stellar age distribution at the solar annulus midplane peaks at $\tau \sim 7.5 \gyr$.  The large difference between the black line and gray regions for $\tau > 10 \gyr$ is due to the fact that age errors strongly skew the distribution toward younger, and therefore dynamically colder stars.  In a similar fashion, at $\tau  = 4 \gyr$ the errors introduce a bias toward older stars, causing the gray band of possible observed AVRs to fall above the intrinsic AVR.

As far as we can ascertain, \SIM\ is the only cosmological zoom-in simulation with a present-day AVR consistent with the MW at the level found here across $13 \gyr$ of evolution.

\subsection{The origin of the AVR}
\label{sec:AVR}

\begin{figure}[!tbp]
  {\centering \includegraphics[width=\linewidth]{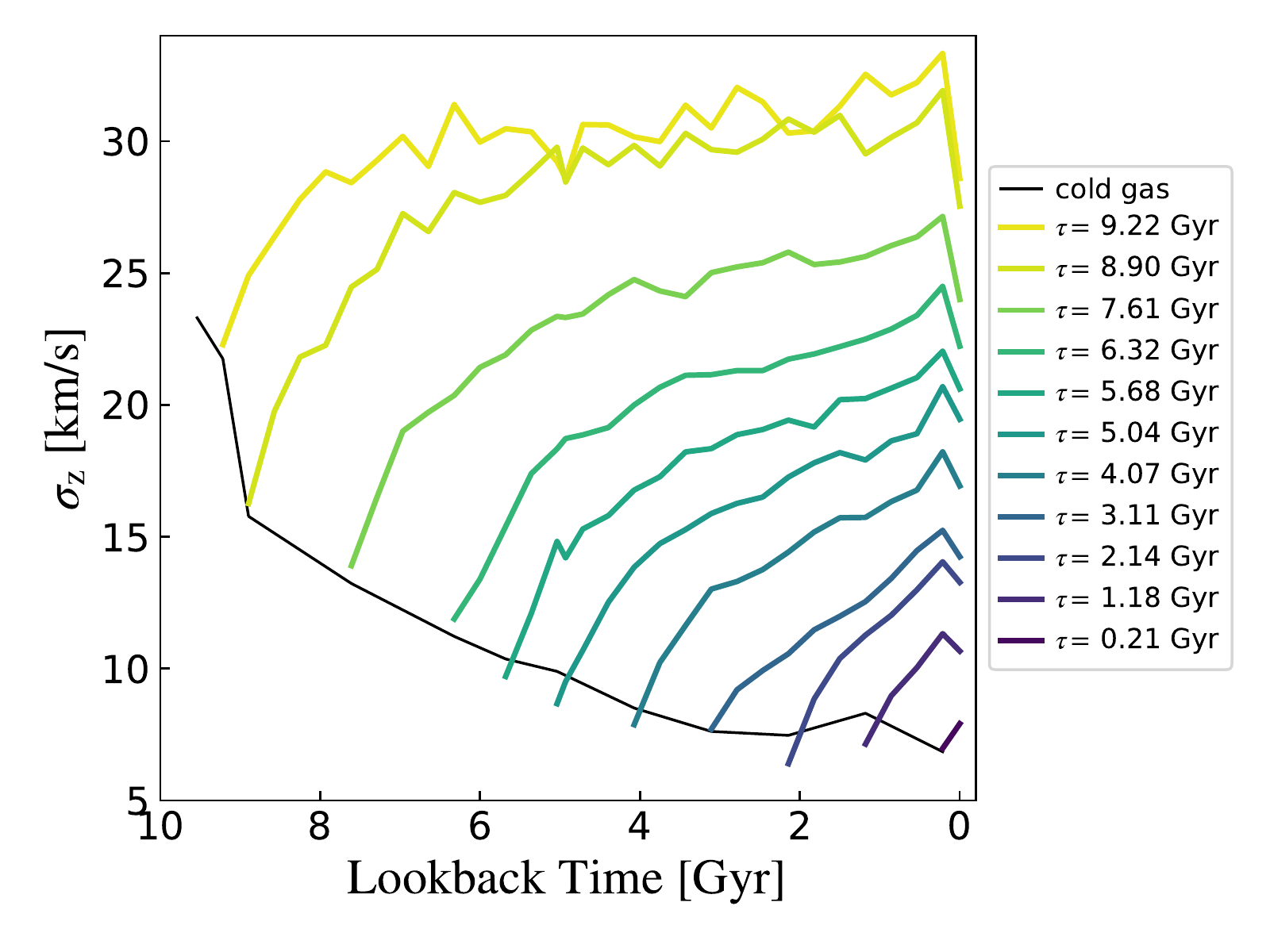}}
  \caption{\label{fig:ud_solneigh}
   The temporal evolution of the vertical velocity dispersion, \vzdisp, for coeval stellar populations (or mono-age populations, MAPs) found in the present-day solar annulus (colored lines) and the star-forming gas reservoir (black line). Line color corresponds to MAP age; darker colors represent increasingly younger stellar populations (see legend).
   The beginning of each line corresponds to the lookback time and \vzdisp\ when a MAP formed. MAPs are born with the kinematics of the gas reservoir at the time of their birth. 
   Both ``Upside-Down'' formation and dynamical heating play a significant role in creating the present-day AVR.
   Older stellar populations are born dynamically hotter than younger ones, and \vzdisp\ increases for all MAPs following their birth.}
\end{figure}

We now investigate how the present-day AVR came to be. Figure~\ref{fig:ud_solneigh} shows the temporal evolution of \vzdisp\ for several stellar cohorts that end the simulation in the solar annulus.
These cohorts are mono-age populations (MAPs); they contain stars born within a narrow range of formation time ($\Delta$\tform = $50$ \myr).\footnote{We confirm that the measured \vbirth\ does change for smaller values of  $\Delta$\tform, see Section \ref{heating_other}}
The end points of each evolutionary track represent the present-day kinematics of each MAP, as encapsulated by the AVR discussed above.
In slight contrast to Figure~\ref{fig:avr}, we use all stars in the solar annulus, regardless of vertical position, as height restrictions would introduce phase correlations that can create a biased view of the velocity dispersion histories \citep{Aumer2016}.  We track the kinematic evolution of both the stellar MAPs (color-coded lines) and the cold ($<1000 \mathrm{K}$) gas  (thin black line) as function of lookback time (\timelb).

The kinematics of the star-forming gas reservoir regulate the assembly of the disk. At each formation epoch, we calculate the \vzdisp\ of the star-forming gas reservoir using the cold gas ($T <1000  \K$) that is spatially coincident with the MAP at birth; i.e., is within the annulus defined by the inner quartile range of formation radius of each  MAP.  The in-situ \vzdisp\ of MAPs at birth are nearly equal to that of the cold gas. Stars clearly inherit the kinematics of the gas from which they form. The star-forming gas reservoir collapses and becomes dynamically colder over time; \vzdisp\ decreases from $\sim 20 \kms$ nearly $10 \gyr$ ago to $7 \kms$ in the present-day.
In turn, progressively younger MAPs form with decreasing \vzdisp.  \citet{Bird2013} called this behavior ``Upside-Down'' formation.  It has been recognized in a wide range of recent cosmological zoom-in simulations  \citep[\eg][]{Grand2016, Ma2017}, but its role in disk kinematics over the last $10 \gyr$ is still debated (see Section~\ref{sec:discussion}). Figure~\ref{fig:ud_solneigh} qualitatively resembles Fig. 19 of \citet{Bird2013}, but in this simulation the birth values of \sigz\ are lower, especially at late times, and growth of \sigz\ by heating plays a larger role (see section~\ref{heating_other}).
The star-forming gas \vzdisp\ decreases in an exponential fashion, dropping by a factor of $2$ from $\timelb\sim10 \gyr$ to $6 \gyr$ but remains nearly constant ($\sim 7 \kms$) for $\timelb < 4 \gyr$.
Upside-Down formation  establishes a correlation between the formation time of a stellar population and its \vzdisp\ at birth due to the evolving kinematics of star-forming gas reservoir.

Stellar populations experience dynamical heating after birth; \ie\ the present-day velocity dispersion is greater than \vbirth\ for each MAP (Figure \ref{fig:ud_solneigh}).
The heating rate $d\vzdisp / dt$ generally appears velocity-dependent; for each MAP it is highest just after formation and decreases with increasing \vzdisp.
Gravitational scattering of the stars due to encounters with a perturbing population, such as midplane concentrated molecular clouds, would produce the general kinematic evolution seen here \citep{Spitzer1951, Lacey1984}.  Indeed, the shape of the $\vzdisp(t_{lookback})$ tracks are qualitatively similar to those of MAPs in idealized dynamical heating experiments \citep[\eg][]{Aumer2016}.  We will investigate the details of the dynamical heating mechanism(s) in future work. Ultimately, it is clear that \SIM's present-day kinematics result from the detailed balance of the collapsing star-forming gas reservoir and the dynamical heating experienced by MAPs after birth.

\subsection{Formation Radius and \vbirth}
\label{sec:32}

\begin{figure}[t]
  {\centering \includegraphics[width=\linewidth]{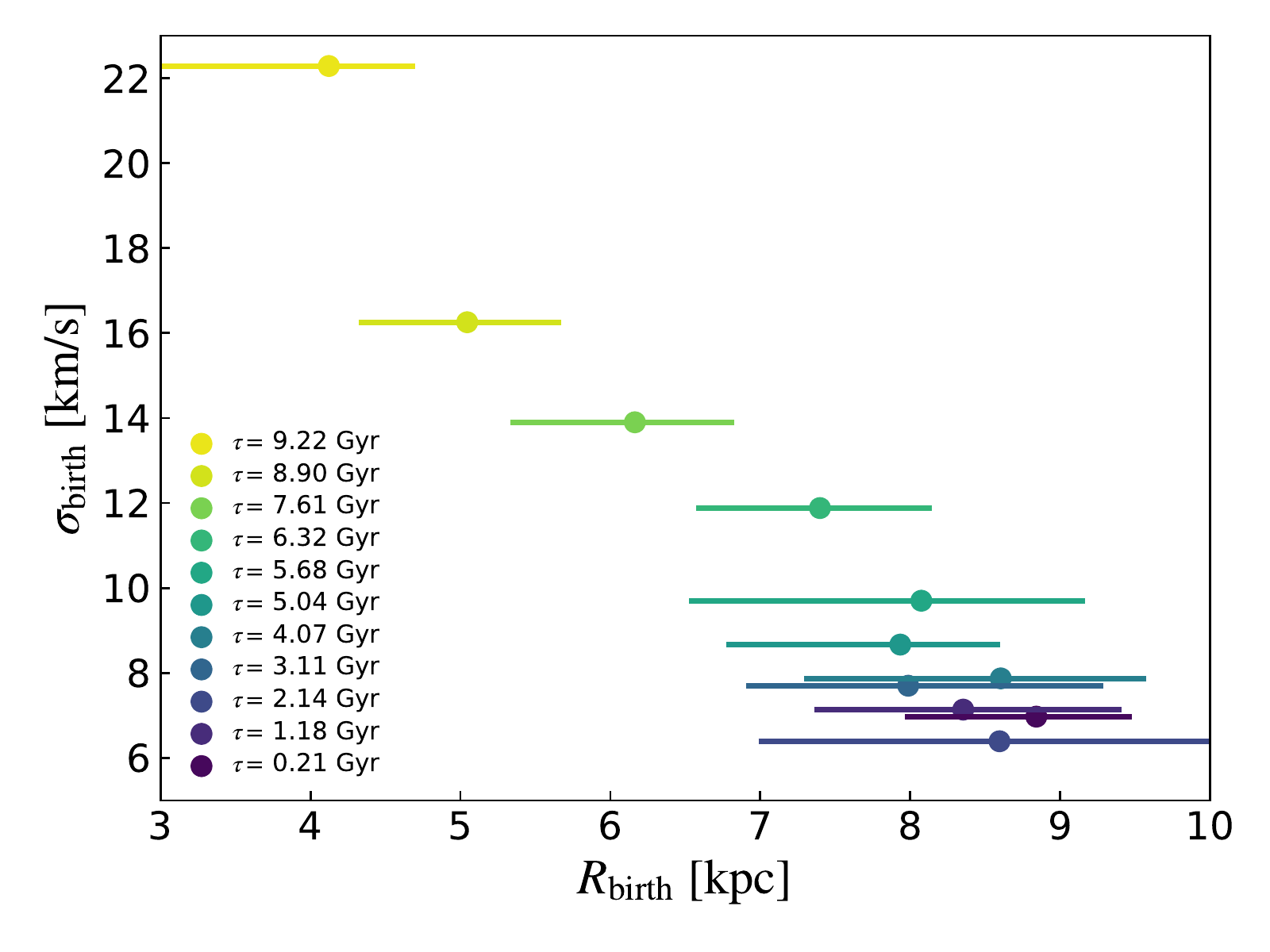}}
  \caption{\label{fig:disp_rform}
  The relationship between vertical kinematics at birth and formation radius of MAPs found in the present-day solar annulus. 
  Points represent a MAP median formation radius and the velocity dispersion at birth; error-bars indicate the interquartile range of \rform.
  MAPs with earlier formation times (color-coding is identical to Figure~\ref{fig:ud_solneigh}) were generally born at smaller formation radii and larger \vbirth\ than their younger counterparts.}
\end{figure}

Many of the stars currently at the solar cylinder were not born there.
Figure~\ref{fig:disp_rform} shows that older stellar populations that end the simulation in the solar annulus were born at smaller radii, \rform, compared to their younger counterparts.  The figure shows the velocity dispersion of the MAP at birth, \vbirth, as a function of \rform. There is a clear trend for stars in the present-day solar neighborhood: progressively younger MAPs were born at larger \rform\ and smaller \vbirth.

This is expected for two reasons.
First, the trend in age versus birth radius seen in Figure~\ref{fig:disp_rform} is a clear sign of inside-out disk growth in \SIM.  The older stars formed in a more centrally concentrated disk.  Very few 10 Gyr old stars were born in the solar annulus.  The vast majority were born in the interior of the disk but are found in the solar annulus at $z=0$. Second, the decreasing trend in \vbirth\ with decreasing age is indicative of disk settling.  The oldest stars considered here have a small median formation radius and large birth velocity dispersion ($\rform, \vbirth$) = ($\sim 4 \kpc,\ 22 \kms$).
At the other extreme, the youngest stars born in the solar annulus have the largest formation radii and are amongst the dynamically coldest at birth ($\rform, \vbirth$) = ($\sim 9 \kpc,\ 7 \kms$). 

Figure~\ref{fig:disp_rform} shows that birth radius and birth kinematics are correlated. Stars born at smaller (larger) galactocentric radii are kinematically hotter (colder) at birth. The inside-out growth of the disk is directly coupled to the disk cooling seen in Upside-Down evolution.
Since age and formation radius are correlated, birth velocity dispersion is also strongly correlated with formation radius as the galaxy's disk cools; i.e., as the galaxy grew outwards, it also grew kinematically cooler.  Thus, the \vbirth\ of a MAP can be a function of both $\rform$ as well as age.

\subsection{SFR density and \vbirth }\label{sec:eqbm}
\label{sec:33}

\begin{figure}[!tbp]
 {\centering \includegraphics[width=\linewidth]{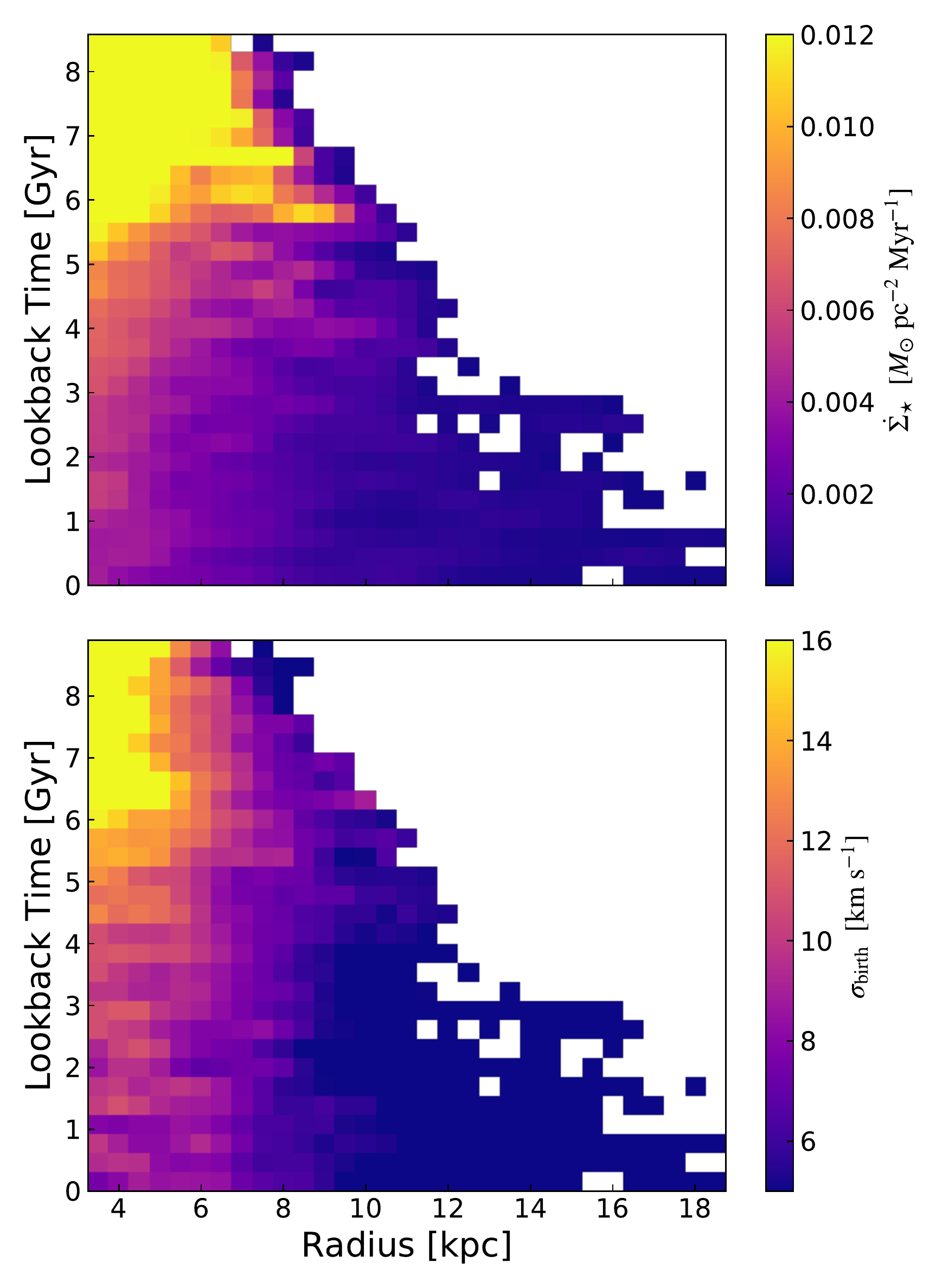} }
  \caption{\label{fig:radprofs} The  star formation rate surface density (top) and the velocity dispersion of stars at birth (bottom) as a function of radius and time.
  We independently smooth each radial profile to emphasize trends; pixel color corresponds to the running mean value across $1.5 \kpc$ in radius ($3$ pixels).
  At a given epoch (row), \sfrden\ is typically highest in the inner galaxy and gradually decreases as the galactocentric radius increases. 
  At a fixed radius (column), \sfrden\ is generally highest at early times and decreases towards the present-day. 
  The radial and temporal gradients in \vbirth\ are very similar.
  Both \vbirth\ and \sfrden\ are strongly correlated at the local spatial scale over much of the galaxy's history. }
\end{figure}

In the bottom panel of Figure \ref{fig:radprofs}, we show how \vbirth\ varies as both a function of formation position and epoch in the disk. Several patterns emerge. First, the typical \vbirth\ in the disk increases with lookback time; \ie\ ``Upside-Down'' formation occurs across the entire disk.
Second, \vbirth\ has a negative large-scale radial gradient over the last $\sim 10 \gyr$; \ie\ the outer disk is dynamically colder than the inner disk at all times.  Thus, the same trends that were seen for solar annulus MAPs also hold for the entire disk. 

Figure~\ref{fig:radprofs} (bottom panel) also shows that the most dramatic evolution of \vbirth\ occurs when combining both the temporal and radial gradients, moving from the upper left to the bottom right of the panels.
The greatest discrepancy in \vbirth\ is found when comparing older stellar populations born in the inner disk with young stars that formed in the outer disk.  These trends manifest themselves in the AVR of the solar neighborhood.

To make these radial profiles, we divide the disk into a grid in $(R, \theta)$. Radial annuli are spaced $0.5 \kpc$ apart and further divided into $18$ azimuthal bins ($\Delta\theta=20\degr$). The velocity dispersion profiles shown here are azimuthally-averaged within each annulus. In general, these results are consistent with what one would find by binning in radius only. In the outer disk, however, warps from either satellite encounters (generally not applicable for \SIM\ over the last $10\gyr$; Section~\ref{sec:sim}) or gas accretion \citep{Roskar2010} can change the mean velocity of gas and stars on one side of the galaxy versus the other, which would artificially increase the measured dispersion if the annulus was taken as a whole. Furthermore, the Toomre stability criterion and many forms of potential pressure support to the ISM act \emph{locally}; if the gas is not well-mixed within an annulus or experiences a much different gravitational potential (more likely in the outer galaxy due to the significantly longer dynamical time), then the velocity dispersion should be calculated on more local scales.

The top panel of Figure~\ref{fig:radprofs} shows that the surface density of star formation, \sfrden, varies with radius and lookback time in the same qualitative fashion as \vbirth.
At a specific lookback time, the inner disk has higher \sfrden\ than the outer disk, and \sfrden\ generally decreases towards the present-day. 
We find that the relationship between \sfrden$(R, \timelb)$ and \vbirth$(R, \timelb)$ is strongly correlated; this is visible in the nearly one-to-one color correspondence between \sfrden\ and $\sigma_{\rm birth}$.
While the dynamic range of each panel image is different, 
the color scale of each panel is linear, and the different dynamic ranges of each panel are suggestive of the slope of the relationship. 
Strong correlation between a gaseous component velocity dispersion and \sfrden\ is expected on theoretical grounds when combining the Toomre stability criterion and the Kennicut-Schmidt law (discussed further below) and has been observed in star-forming galaxies \citep[\eg][]{Yu2019}.
We find that \sfrden\ correlates with \vbirth\ at the site of star formation and that this correlation exists throughout the disk and for at least the past $\sim10 \gyr$ of evolution.

\begin{figure}[!tbp]
 {\centering \includegraphics[width=\linewidth]{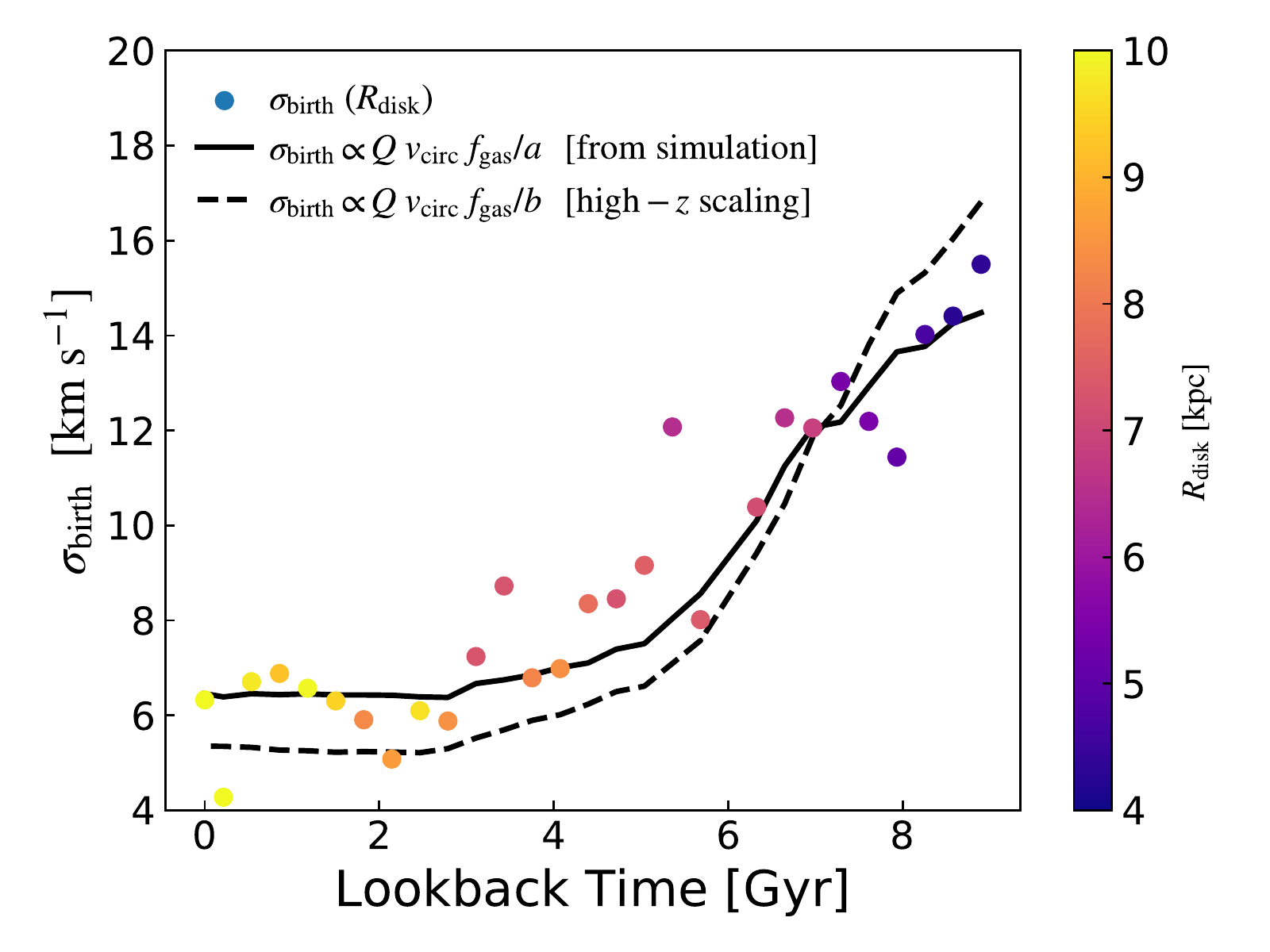}}
  \caption{\label{fig:highz} Comparison of simulation measurements and gravitational equilibrium-driven  predictions for the evolution of \vbirth. Points show the vertical velocity dispersion of newly formed stars in an annulus at \rdisk\ as a function of \timelb, colored by \rdisk. \rdisk\ is the radius that encloses 80\% of ongoing star formation, and \vbirth\ is measured within an annulus at \rdisk\ $\pm$ 0.5 kpc.  \rdisk\ increases towards the present-day as the disk grows inside-out. The two black lines are predictions for the evolution of $\vbirth$ assuming the disk is marginally stable.  See text for details.  Both models predict a disk settling process very similar to what is measured, suggesting that the disk is continuously driven towards marginal gravitational instability, which plays a strong role in governing the slow collapse of the star-forming gas reservoir (and hence \vbirth).
  }
\end{figure}

\subsection{Disk Stability}
\label{sec:highzmodel}

We characterize the evolution of \vbirth\ for the global disk in Figure~\ref{fig:highz}.
Here, we define the radius of the star forming disk, \rdisk, to be the radius that encloses 80\% of the ongoing star formation.  We measure \vbirth\ in an annulus at \rdisk$\pm 0.5 \kpc$. 
The upside-down formation of the disk is clearly seen as \vbirth\ decreases towards the present day (colored points). 
The velocity dispersion of stars at birth decreases by a factor of $\sim2.5$ over the last $10\gyr$, but much of that evolution occurred early and \vbirth$(\rdisk)$ remains below $8 \kms$ over the past $5 \gyr$.
The disk also grows outward during this time; \rdisk\ steadily increases as star-formation proceeds further out in the in disk \citep[represented by the colorbar in Figure~\ref{fig:highz}; see also][]{Minchev2013}. 
Empirically, as \SIM\ grows inside-out, it also grows upside-down.

The kinematic evolution of \vbirth\ is in good agreement with scaling arguments under the assumption that the disk is in equilibrium, \ie, that the disk remains quasi-stable as new gas accretes onto the galaxy to replace gas lost to star-formation and outflows \citep[\eg][]{Thompson2005, Krumholz2012, Dave2012, Dekel2014}.
If the gas component provides the dominant component to the midplane pressure, the Toomre relation \citep{Toomre1964} of the star-forming gas reservoir can be re-written as 
\begin{equation}
\vbirth / v_{\mathrm{circ}} = \psi \fgas Q / \sqrt{2}
\label{eq:toomreQ}
\end{equation}
where $v_{\mathrm{circ}}$ is the rotation speed, $Q$ is the Toomre parameter, $\psi$ is a constant of order unity, and \fgas\ is the gas fraction \citep{Genzel2011}.
The $\sqrt{2}$ factor stems from the epicyclic frequency component of the Toomre relation and our assumption of a flat, or nearly flat, rotation curve.
The quantity \fgas\ is the fraction of the overall mass in gas that contributes to the effective gravitational potential felt by the newly-forming stars.  

We next demonstrate that \vbirth$(\rdisk)$ can be reproduced assuming a Toomre stable disk, with \vbirth\ being determined by the evolving gas fraction \fgas.   The circular velocity at \rdisk\ is nearly constant, varying less than $\sim10\%$, over the past $10 \gyr$. The lines in Figure \ref{fig:highz} are model predictions that assume a constant $Q_{\mathrm{gas}}=1$ at \rdisk\ for all epochs.  We compute \fgas\ in two ways.
In the first model, \fgas\ is calculated using the empirical scaling \citep{Tacconi2013}
\begin{equation}
    \fgas = \frac{1}{1 + (\tau_{\mathrm{depl}}\ \mathrm{sSFR})^{-1}}
\label{eq:fgas_emp}
\end{equation}
where the specific star formation rate sSFR$=\dot{M_{\star}} / M_{\star}$ and $\dot{M_{\star}}$  is measured directly from the simulation as the average SFR over the last $50 \myr$ and the gas depletion time is modeled as $\tau_{\mathrm{depl}} = 1.5 \times (1 + z)^{\alpha} \gyr$  with $\alpha=-1.0$.\footnote{$\alpha$ has been measured between $-0.7$ and $-1.5$ \citep{Dave2012, Tacconi2013}}
Many empirical studies of resolved spectroscopy of disk galaxies use Equation \ref{eq:fgas_emp} \citep[\eg][]{Wisnioski2015}.
Second, we compute the fraction of the gravitational potential due to the gas surface density,
\begin{equation}
    \fgas = \frac{\Sigma_{\mathrm{gas}}}{\Sigma_{\mathrm{tot}}} \ \bigg\vert_{\rdisk}
    \label{eq:fgas_Sigma}
\end{equation}
where $\Sigma_{\mathrm{tot}}(R)= v_{\mathrm{circ}}^2 / (\pi G R)$ is the total dynamical surface density.
The gas surface density is not for all gas, just the fraction of the ISM that contributes to the vertical midplane pressure felt by the newly forming stars \citep[\eg][]{Thompson2005, Krumholz2012}.
To best track the mass of the ISM directly contributing to star formation, we select all gas with a density higher than $1\, \mathrm{atoms}\ \mathrm{cm}^{-3}$. Here, \fgas\ is evaluated at \rdisk.

We can rewrite Equation \ref{eq:toomreQ} as $\vbirth \propto v_{\mathrm{circ}} \fgas Q / \mathrm{C}$, where C is a constant chosen to match the normalization of \vbirth$(t)$.  Figure \ref{fig:highz} shows $\rm{C}=2$ when \fgas\ is set by the empirical high-$z$ scaling (dashed line), and  $\rm{C}=0.4$ when \fgas\ is set by the dense gas surface density in the simulation (solid line). The evolving gas fraction largely determines the shape of both lines as \vcirc\ changes by only $\sim10\%$, and the Toomre $Q$ parameter is assumed constant ($Q=1$).  

Note that the recast marginal stability argument is \emph{not} a fit to the $\vbirth$ measurements, yet it predicts a disk settling process very similar to what is measured.  This suggests that the disk is continuously driven towards marginal gravitational instability, which plays a strong role in governing the slow collapse of the star-forming gas reservoir (and hence \vbirth).

\section{Discussion}
\label{sec:discussion}

In this section we first discuss the key physical processes that occur in the simulated galaxy that result in the excellent match to the Milky Way's AVR at $z=0$.  We then compare to previous simulation work, and highlight the key numerical achievements that have allowed these physical processes to occur.

\subsection{Matching the Milky Way's AVR}
\label{sec:how}

We find (Figure \ref{fig:avr}) that the solar neighborhood AVR measured in \SIM\ is nearly identical to the AVR in the solar neighborhood of the MW.  This is true for stellar populations that span more than 10 Gyr in age. The primary physical process that allows \SIM\ to match the AVR over such a large age range is disk settling, or upside-down formation, combined with inside-out disk growth.  We discuss below that the resulting \vbirth($z$) is set by the fact that the stars in the disk are born in equilibrium.  Figure \ref{fig:ud_solneigh} also shows that stellar populations are subsequently heated after birth, and that the velocity dispersion of the cold gas in \SIM\ over the last few Gyr prior to $z=0$ is an excellent match to the observed $\sigma$ in star forming gas near the Sun.  Below, we discuss these factors that set the AVR.

\subsubsection{Stars are Born in Equilibrium}

We have shown that \SIM\ exhibits both inside-out and upside-down formation of disk stars.  The inside-out growth of galaxy disks is expected due to the fact that the angular momentum of accreted gas increases with decreasing $z$ \citep[e.g.,][]{White1978,Fall1980}.  Inside-out growth has been observationally confirmed in relatively massive disk galaxies \citep[e.g.,][]{Williams2009, Gogarten2010, Barnes2014, Frankel2019, Peterken2020}.  The theoretical explanation behind the upside-down formation of disks has only more recently been explored, motivated by the observations of kinematically hot disks at high $z$.  It has been shown that this evolution can be modeled if the velocity dispersion of gas evolves according to the gas fraction within a marginally stable disk \citep[e.g.,][]{Wisnioski2015}.

We find (section \ref{sec:eqbm}) that the velocity dispersion of newly forming stars, \vbirth, at any radial position in the disk correlates with the star formation rate surface density, \sfrden, at that radius over the past 10 Gyr (see Figure \ref{fig:radprofs}).  This behavior is expected if a disk is marginally stable, in dynamical equilibrium, and follows a Schmidt-Kennicutt relation \citep{Schmidt,Kennicutt1998} for star formation \citep[\eg][]{Krumholz2012, Swinbank2012}.  The star formation prescription adopted for this simulation ensures it follows a Schmidt-Kennicutt relation \citep{Christensen2012, Christensen2014a}.  
It is also clear from Figure \ref{fig:ud_solneigh}  that \vbirth\ tracks the velocity dispersion of the star forming gas, $\sigma_{\mathrm{coldgas}}$, and $\sigma_{\mathrm{coldgas}}$ should track the neutral gas velocity dispersion $\sigma_{\mathrm{g}}$ to within a factor of two.  If the disk has a constant $Q_{\mathrm{gas}}$, $\sigma_{\mathrm{g}} \propto \Sigma_{\mathrm{g}}$, where $\Sigma_{\mathrm{g}}$ is the surface density of the neutral gas. 
Meanwhile, the Schmidt-Kennicutt relation requires \sfrden\ $\propto$ $\Sigma_{\mathrm{g}}^{1.5}$ \citep{Leroy2008}\footnote{There is substantial scatter in the Schmidt-Kennicutt relation at small scales \citep{Christensen2012, Orr2019} }. Putting it all together, $\vbirth \propto \sigma_{\mathrm{g}} \propto \Sigma_{\mathrm{g}} \propto$ \sfrden$^{1.5}$\ for a stable disk in equilibrium. Using the data from Figure~\ref{fig:radprofs}, we find $\vbirth \propto \sfrden^{\sim0.4}$. This results suggests that $\vbirth$ is governed by something more than a constant $Q_{\mathrm{gas}}$.  

On the other hand, Figure \ref{fig:highz} demonstrates that a dynamical equilibrium model can reproduce the trends seen in \SIM's stellar populations at \rdisk.  For a moderately stable gas disk in equilibrium, the evolution shown in Figure \ref{fig:highz} is driven by the decline in gas fraction, $f_{\rm gas}$, with decreasing $z$.  Likewise, the evolution of the galactic SFR in \SIM\ nearly matches the qualitative evolution in \vbirth$(\rdisk)$ seen in Figure \ref{fig:highz}.  The SFR of \SIM\ was a factor of $\sim 4-5$ higher $\sim10 \gyr$ ago relative to the current SFR ($\sim 1 \msol / {\rm yr} $), decreased relatively quickly to $1.5 \msol / {\rm yr}$  nearly $5 \gyr$ ago, and showed little change over the past $4 \gyr$.  Because we expect that the SFR is tied to the abundance of fuel for star formation, i.e., $f_{\rm gas}$, it is not surprising that the SFR follows a similar trend.
The star formation history of \SIM\ over the last $10 \gyr$ appears to be fairly consistent with star formation history (SFH) of  the Milky Way \citep[\eg][]{Snaith2015}, which is likely another reason that the AVR of \SIM\ is a good match to the Milky Way's AVR.

What component(s) of the disk are marginally stable? The results in Figures \ref{fig:radprofs} and \ref{fig:highz} appear to be in tension at first glance.  If $\vbirth \propto \sfrden^{\sim0.4}$ from Figure \ref{fig:radprofs}, it suggests that $Q_{\mathrm{gas}}$ is not constant, and yet a constant $Q_{\mathrm{gas}} = 1$ appears to match the evolution in Figure \ref{fig:highz}. A resolution can be had if the  multi-component $Q_{\mathrm{tot}}$ \citep[\eg][]{Romeo2011}, accounting for contributions from both the gas and stars, is instead the conserved quantity and if $Q_{\mathrm{gas}} \propto Q_{\mathrm{tot}}$ at \rdisk\footnote{The relationship between $Q_{\mathrm{gas}}$ and $Q_{\mathrm{tot}}$ need not be linear away from \rdisk}. Cursory investigation suggests that $Q_{\mathrm{tot}}(R, \tau)$ is the quantity that is relatively conserved rather than $Q_{\mathrm{gas}}$ across the disk and that the gas and stellar components contribute approximately equally to local stability at \rdisk. We will investigate the multi-component stability of the disk in detail in future work. Resolved studies of local star-forming disk galaxies \citep{Romeo2013} and their simulated counterparts \citep{Orr2019} also connect the kinematics of star-forming gas with multi-component stability. 
Conservatively, we conclude a quasi-constant $Q_{\mathrm{tot}}$ is plausible and consistent with all of the results in this manuscript. 

Vertical hydrostatic equilibrium is a common assumption in analysis of the ISM, and controlled, high-resolution simulations show that equilibrium is quickly established even with a wide range of initial conditions \citep[\eg][]{Koyama2009, Kim2013}. Controlled simulation experiments have shown that galaxies naturally regulate their star formation, striking a balance between the pressure support in the midplane and the weight of the ISM, even when varying sub-grid physics \citep[\eg][]{Benincasa2016}. 
In future work, we will determine the local gravitational stability of the disk and test the reasonable assumption that stars are born from gas in hydrostatic equilibrium with the local disk.

\subsubsection{Heating/Radial Mixing}

The formation radius of the stars has a big role to play in \vbirth, not just the time at which they formed. This is because the galaxy is not only forming upside-down, but inside-out.
The effective radius of the disk (\rdisk, the mass or light-weighted mean radius of newly formed stars) moves outwards as the galaxy grows (Figure~\ref{fig:highz}).  As we have just established, older stars are born at a time when \vbirth\ is higher, likely due to the higher $f_{\rm gas}$ of the disk while in marginally stable equilibrium, and a higher SFR.  For \SIM, the oldest stars contributing to the AVR were also born at roughly half of the radius of the solar neighborhood.  Younger stars are born with lower \vbirth, and closer to the solar annulus. 

As can be seen in Figure \ref{fig:radprofs}, even at $z=0$ the inner disk has a higher \vbirth.  Therefore, both a smaller formation radius and older age contribute to higher \vbirth. 
As can be read from the bottom panel of Figure \ref{fig:radprofs} (upper left to lower right), this evolution leads to a maximizing of the evolution of \vbirth\ for stars that end up in the solar neighborhood at $z=0$. 
The radial dependence of \vbirth\ increases the difficulty of inferring the in-situ birth kinematics of a stellar population given present-day observations. The dynamic range and distribution of \vbirth\ for stars at any present-day radius is influenced by both \vbirth$(\tau)$ and the history of stellar radial mixing away from \rform\ specific to the galaxy of interest. How the stars migrated away from their birth radii, whether by changes in orbital angular momentum \citep[\eg][]{Sellwood2002} or simple increases in orbital energy due to gravitational scattering, does not alter this conclusion. 

It is clear from Figure \ref{fig:ud_solneigh} that stellar populations in the solar neighborhood at $z=0$ have been heated after birth. In-situ kinematics at birth account for nearly half of the present-day velocity dispersion of each solar neighborhood MAP ($\vbirth / \vzdisp \sim 0.4$, Figure~\ref{fig:ud_solneigh})\footnote{In fact, there is a small range in $\vbirth / \vzdisp$ $\sim[0.4,0.5]$, with the oldest stars having the lowest ratio and younger stars having the highest ratio}. The source of the heating is likely responsible for the radial mixing as well.  Although mergers have been shown to heat the disk \citep{Martig2014, Grand2016, Buck2020}, \SIM\ has a very quiescent merger history over the past 10 Gyr, so mergers are unlikely to be a major source of heating in this particular galaxy.  We note that \citet{Grand2016} found that a central bar is one of the strongest factors in radial mixing, and \SIM\ had a bar in its past. Also, as we discuss further below, \SIM\ is capable of resolving the cold, clumpy structure of the ISM.  It has long been suggested that scattering of stars off molecular clouds could heat a stellar population as it ages \citep{Spitzer1951}.  The fact that the heating rate of the MAPs depends on velocity (Figure \ref{fig:ud_solneigh}) strongly suggests that GMC scattering is occurring in \SIM. 

\citet{Leaman2017} used a dwarf galaxy simulation with identical physics to \SIM\ to investigate the AVR in the Wolf-Lundmark-Melotte (WLM) galaxy. They assumed that \vbirth\ was tied to the SFH (and hence $f_{gas}$) of WLM, again adopting an equilibrium model. They showed that additional GMC heating after birth could fully explain the stellar dynamics of the simulated dwarf, which was an excellent match to the AVR of WLM.  However, they concluded that \vbirth\ plus GMC heating alone is insufficient to explain the AVR in more massive disk galaxies like the MW, and that additional heating by mergers or bars is necessary. However, \citet{Leaman2017} did not consider the impact of \rform\ on \vbirth. Our results suggest that the \vbirth\ of old stars in the present-day solar neighborhood is likely enhanced over the global \vbirth$(\tau)$ prediction due to their small \rform. If so, additional heating from mergers would be unnecessary for stars younger than $\sim10 \gyr$. In contrast, there have been recent claims that the detailed AVR of the MW (for $\tau \lesssim 8 \gyr$) can arise solely from steady dynamical heating from giant molecular clouds and, optionally, spiral waves, without the need for an evolving \vbirth\ \citep{Ting2018, Mackereth2019}.

We note that it is nearly impossible, based only on the present-day vertical kinematics of these stars, to distinguish between a pure dynamical heating evolutionary history and disk settling combined with heating. The evolution of vertical energy ($E_z \sim \varz$) due to gravitational orbit scattering is $\mathrm{d}E_z/\mathrm{d}t \sim \Sigma_{\mathrm{pert}} / E_z$, where $\Sigma_{\mathrm{pert}}$ is the surface density of the orbit perturbers \citep{Binney2008}. The $E_z$ dependence of the heating rate means that there is little difference, at a fixed $\Sigma_{\mathrm{pert}}$ (typically the surface density of molecular clouds), in the present-day vertical energy of, \eg\ a $\sim 7 \gyr$ population with $\vbirth= 15\kms$ vs $7 \kms$.  This is because any increase in \vbirth\ is largely negated by a decreased heating rate.  However, there is hope to distinguish between these evolutionary tracks using the shape of the velocity ellipsoid, $\sigz/\sigma_{r}$, \citep[see, e.g.,][]{Mackereth2019}, which is more sensitive to the particular heating history \citep[\eg][]{Jenkins1990, Lacey1984, Ida1993, Sellwood2008}.  We will examine the connection between the velocity ellipsoid and the kinematic history of \SIM's solar neighborhood stars in a future work.

\subsection{Comparison to Previous Work}
\label{sec:why}

As we have just discussed, it is the fact that \vbirth\ decreases with decreasing redshift, combined with the fact that the disk is growing inside-out, that is the primary origin of the solar neighborhood AVR in \SIM. Subsequent heating also plays a role. Below we compare our results from \SIM\ to previous simulation results, and establish that it is the simulation's ability to produce dynamically cold gas that allows both the upside-down and inside-out growth to emerge.

\subsubsection{Dynamically Cold Gas}

To the best of our knowledge, this simulation is the first cosmological zoom-in MW-mass simulation to recover the low velocity dispersions of young stars in the solar neighborhood at $z=0$.  In \SIM, stars forming in the solar neighborhood over the last 3-4 Gyr before $z=0$ are born with exceptionally cold dynamics, with a velocity dispersion as low as $\sim 6 \kms$. This velocity dispersion is consistent with the internal kinematics of nearby molecular clouds in the MW \citep{Stark1989} and the average cloud-to-cloud velocity dispersion of molecular gas in nearby disk galaxies \citep[$6.1\pm1.0 \kms$][]{Wilson2011}. The ability to form stars with this low dispersion is critical to matching the MW AVR, particularly for young stars.

\SIM\ adopts the star formation recipe from \citet{Christensen2012}, which explicitly follows the non-equilibrium formation and destruction of H$_2$, and restricts star formation to occurring only in the presence of H$_2$. This restriction ensures that stars are born from a much dynamically colder star-forming gas reservoir, with low temperatures and at high densities ($T < 1000$ K, $n > 100$ atoms cm$^{-3}$).  

The non-equilibrium H$_2$ recipe adopted here seems to be the key to setting a low \vbirth\ at the present day.  \citet{House2011} examine how \vbirth\ varies with star formation prescription, and find that a dispersion ``floor'' is created if cold gas is not resolved \citep[see also][]{Martig2014}.  Likewise, \citet{Kumamoto2017} show that as the threshold for star formation is decreased, \vbirth\ increases because the star forming region becomes ``thicker.''  

At higher star formation thresholds, and thus higher \vbirth, upside-down disk settling cannot occur because the disk is born hot.  This is seen in earlier works \citep{House2011, Martig2014} in which quiescent disks are born with a constant (high) \vbirth\ over the last 9-10 Gyr instead of cooling with time.  Likewise, if the cold, clumpy ISM isn't resolved, then subsequent heating due to GMC scattering is weakened.  Even if GMC scattering is present, the heating rate is diminished if the stellar population is already born hotter, as discussed above.  Thus, it is the ability of \SIM\ to capture this dynamically cold gas that allows the primary processes that set the AVR to emerge.  

\citet{Kumamoto2017} studied the evolution of \vbirth\ in isolated, non-cosmological disk galaxies, and found results very similar to those in \SIM.  With their high resolution, they were able to form stars at similar densities as in \SIM\ ($n > 100$ atoms cm$^{-3}$), but even colder temperatures ($T < 100$ K).  The stars in their simulation have \vbirth\ as low as 3 \kms.  As in \SIM, \citet{Kumamoto2017} find that \vbirth\ decreases in their simulated galaxy as $f_{\rm gas}$ decreases.  Because they are able to resolve the clumpy ISM, they also find that stellar populations were heated after birth due to GMC scattering, and that the rate of heating decreases with time as the amount of dense gas decreases. 

\SIM\ appears to be the first cosmologically simulated disk galaxy to reproduce the low $\sigma$ values observed in GMCs, allowing it to match the observed AVR in the MW's solar neighborhood.  The results of \citet{Kumamoto2017} may suggest that \vbirth\ could be even lower if lower temperatures could be used for star forming gas.  On the other hand, observationally \vbirth\ seems to be $\sim$6 \kms\ \citep[e.g.,][]{Stark1989, Wilson2011} rather than the 3 \kms\ found in \citet{Kumamoto2017}, so it is possible that something acts to keep the dispersion ``floor'' closer to 6 \kms\ in real galaxies.

\subsubsection{The Role of Feedback}

\SIM\ is not the first cosmological disk galaxy simulation to form stars at high densities and low temperatures, yet it seems to be the first with such a low \vbirth.  For example, \citet{Sanderson2020} show the AVR of simulated MW-mass disk galaxies from the FIRE-2 suite, which forms star particles in self-gravitating molecular gas that can self-shield, at $n > 1000$ atoms cm$^{-3}$.  Despite this high star formation threshold, the youngest stars in their simulations have a (total) velocity dispersion at least 20 \kms\ higher than young stars in the MW.  This suggests that some other factor may influence \vbirth\ in simulations, but that this factor is not as dominant in \SIM.  

Both observationally and theoretically, it has been suggested that there is some threshold \sfrden\ above which supernovae are the dominant driver of turbulence \citep[e.g.,][]{Agertz2009, Tamburro2009, Stilp2013}.  Stellar feedback and supernovae are implemented very differently in FIRE-2 than in \SIM.  \SIM\ adopts the ``blastwave'' SN feedback recipe of \citet{Stinson2006}.  This model injects only thermal energy from supernovae into neighboring gas particles, and disables cooling in gas particles that are affected by SN\,II energy (SN\,Ia are allowed to immediately cool).  The disabled cooling is designed to mimic the adiabatic expansion phase of a supernova remnant, which is not otherwise resolved by the simulation. In contrast, FIRE-2 does not disable cooling, but implements a host of young star feedback which is not accounted for in \SIM, e.g., photo-ionization by young massive stars, momentum injection from stellar winds, and radiation pressure on dust \citep{Hopkins2018}.  

\citet{ElBadry2016} found that stars in FIRE are able to form in gas that is outflowing due to feedback, which is likely to impart a large \vbirth.  We do not see this same effect in our simulations.  Rather, the low velocity dispersion of the star-forming gas establishes a pressure ``floor,'' and new stars are born with similarly low $\sigma$.  The numerical implementation and sub-grid physics of \SIM\ do not artificially force the star-forming gas to broader velocity distributions.

Given the computational cost of running this galaxy, we have not explored the direct impact of our chosen physics by re-simulating the galaxy with different feedback. However, we can speculate on how the feedback impacts the results given previous work.  \citet{Agertz2013} explored the impact of varying feedback models on the ISM in galaxies.  They found that the model with delayed cooling, similar to our blastwave model, can better reproduce the star formation history of a galaxy relative to a model with more physically motivated feedback.  While it tends to create more hot gas than other models, in the outer disk it predicts a similar gas velocity dispersion.  Thus, it is not immediately clear that our velocity dispersion results would be different with a different choice of feedback model.  On the other hand, the results from FIRE-2 suggest that the choice of feedback model can definitely play a role in \vbirth.

While feedback in \SIM\ may be injecting turbulent energy that affects the gas dispersion, it appears that it is not substantially boosting \vbirth. The low SFR over the last few Gyr prior to $z=0$ may help to set the dispersion at a constant, low level.  We will explore the role of feedback further in future work.  Overall, an equilibrium disk model combined with dynamically cold star formation seems to be the key to setting the AVR normalization in this simulation.

\subsubsection{Other Factors that Impact Heating Results}
\label{heating_other}
The degree to which heating plays a role in shaping the present-day AVR differs slightly from the results of \citet{Bird2013}.  There, the birth velocity dispersion always accounted for more than half of the present-day value for each MAP, $\vbirth / \vzdisp$ $\sim[0.6,0.7]$, with older stars having a higher ratio than younger stars (see their Figure 19).
As discussed above, their different star formation threshold  ($T < 3\times10^4$ K, $n > 5$ atoms cm$^{-3}$) plays some role, but we have identified additional reasons why our results differ.  

\citet{Bird2013} examined a broader swath of the disk ($4 < R < 8 \kpc$).  As we have seen in this work, extending the inner regions to encompass the inner disk will include hotter stellar populations. Here, rather than adopt a fix radius to define the solar neighborhood, we have scaled the solar neighborhood based on the relative difference in scale length $R_d$ of the simulated disk compared to the MW. \citet{Boardman2020} showed that the scatter in various radial properties of disk galaxies is minimized when scaling by $R_d$, suggesting that this scaling is the best way to directly compare to the MW's AVR.

Finally, we believe that the largest source of difference in our results relative to \citet{Bird2013} is due to the fact that Bird et al.\ identified MAPs with a much larger age range. In this work, single stellar populations span a very narrow range in formation time ($\Delta\tform = 50 \myr$); the specific formation time ranges are chosen such that any MAP forms just prior to a recorded simulation output (\ie\ an epoch that was written to disk). The oldest stars included in the \vbirth\ measurements are therefore $50 \myr$ old, at maximum. In contrast, \citet{Bird2013} define cohorts with $\Delta\tform \gtrsim 1 \gyr$. Dynamical heating is typically strongest right after birth (see Figure~\ref{fig:ud_solneigh}).  Thus, the $\Delta\tform$ used in Bird et al.\,is likely to have increased \vbirth\ from its true in-situ value. We note that the main results of \citet{Bird2013} are not changed by this realization; the in-situ kinematics of stars at birth still evolved with formation time and had a significant impact on the present-day kinematics of each cohort.

To ensure that the present analysis was not affected in a similar manner, we experimented with even smaller ranges of formation time and found no decrease in \vbirth\ for populations as young as $0.5 \myr$. We are therefore confident that the \vbirth\ values reported here truly reflect the in-situ kinematics of each MAP, and that we have accurately captured the subsequent heating of these populations soon after birth.

\subsection{Connection to High $z$ Disks}

The equilibrium disk model that we have shown reproduces the evolution of \vbirth\ for stars in \SIM\ (Figure \ref{fig:highz}) has also been shown to reproduce the AVR in Local Group galaxies \citep{Leaman2017}, as well as the evolution in $\sigma(z)$ seen in disk galaxies \citep[e.g.,][]{Swinbank2012, Wisnioski2015, Turner2017, Ubler2019}.  This model tracks the midplane pressure, which is correlated with \sfrden.  In this model, dependence of \vbirth\ on position (Figure \ref{fig:radprofs}) is expected at late times. Observations of resolved, nearby galaxies find negative radial gradients in both the atomic and molecular gas velocity dispersion \citep{Tamburro2009}.
At a fixed epoch, \vbirth\ should decrease with increasing radius if the midplane surface density decreases with radius and stars are born in hydrostatic equilibrium.  Similarly, \vbirth\ should increase with $z$ as $f_{\rm gas}$ increases, and the kinematics of newly forming stars is thus a function of both position and time within the disk. This not only has important implications for observations of resolved stellar kinematics in the MW and local galaxies \citep{Leaman2017}, but also for the kinematic evolution of disk galaxies out to $z\sim2$.

The observed decrease in the gas velocity dispersion of disk galaxies with decreasing $z$ has been dubbed ``disk settling.'' \citet{Wisnioski2015} demonstrated that disk settling could be explained within a model wherein the velocity dispersion of gas evolves according to the gas fraction within a stable disk. Upside-down growth is consistent with disk settling and implies that it is an evolutionary sequence traversed not just by the population, but by individual galaxies as well.  In fact, \citet{Kassin2014} analyzed \SIM\ (along with three other Milky Way-mass galaxies run with the same physics) and found that the global gas velocity dispersion, $\sigma_g$, in the simulations decreases with time. They found that the slope of the $\sigma_g(z)$ evolution  for the simulated galaxies is consistent with observed evolution, though they were unable to match the normalization of $\sigma_g(z)$ due to the difficulty of identifying gas in the simulations that is a proper match to the star formation-heated gas that observers use as a tracer.  Despite showing that these galaxies match the observed evolution, \citet{Kassin2014} did not explore the origin of the evolution.  The results in this work explain the Kassin et al.\,result in light of the same models applied to explain observed disk settling.  The settling of the disk in \SIM\ appears to be strongly linked to the evolving SFR balanced by the supply of gas for star formation.

\section{Conclusions}
\label{sec:conclusions}

We find the solar neighborhood AVR measured in \SIM\ is nearly identical to that in the MW. To the best of our knowledge, this is the first cosmological zoom-in MW-mass simulation to recover the low velocity dispersions expected of molecular gas at $z=0$, and stars in the simulation are then born with these low  dispersions. The velocity dispersion of the star-forming gas in \SIM\ is $\sim6 \kms$ at $z=0$ in the simulated solar neighborhood (Figure \ref{fig:disp_rform}). Typically, the velocity dispersion of the star-forming gas reservoir stays $>10 \kms$ in cosmological zoom-in simulations, even of MW-like galaxies \citep[\eg][]{Roskar2013, Bird2013, Minchev2013, Martig2014, Sanderson2020, Buck2020}. 
The detailed AVR arose self-consistently from the formation and dynamical evolution of \SIM. It is therefore interesting to analyze how the AVR came about. Through a combination of upside-down and inside-out formation, the resulting AVR yields an excellent match to the MW AVR for stellar populations through $> 10 \gyr$ old.  

There has been a lot of debate about the relative influence of dynamical evolution on stars in the solar neighborhood.  In this work, we examine the kinematics of stars in the simulated solar neighborhood, including at their birth and their subsequent evolution, in order to understand how the present-day AVR of the solar neighborhood came to be. In addition, we consider as a whole the simulated galactic disk at high redshift.  This analysis allows us to draw comparisons between the vertical disk dynamics observed at $z=0$ and observed dynamics in MW proxies observed at high $z$.

Although this is just a single simulated galaxy, \SIM\ has managed to reproduce a number of observed trends in disk galaxy evolution. In this work, we have shown
\begin{itemize}
\item{The radius of the star-forming disk in \SIM\ grows with time (inside-out, Figure \ref{fig:highz}), and \vbirth\ decreases with time.  At any given time, \vbirth\ depends on radius.  At any given radius, \vbirth\ depends on time, consistent with an upside-down formation in which the disk cools or settles with time.  Thus, \vbirth\ is a function of both formation time and formation radius within the disk, and is strongly correlated with \sfrden\ (Figure \ref{fig:radprofs}).}

\item{The oldest stars in the solar neighborhood in \SIM\ were born at smaller radii (Figure \ref{fig:disp_rform}), but contribute to the $z=0$ AVR due to radial mixing.  On the other hand, the youngest stars in the solar neighborhood tend to be born locally.  Given the dependence of \vbirth\ on time and radius, radial mixing tends to amplify the \vbirth\ evolution with age seen in the outer disk and specifically at the solar neighborhood.}

\item{Mono-age stellar populations are also kinematically heated after birth (Figure \ref{fig:ud_solneigh}). While we have not directly explored the origin of the heating in this paper, the ability of this simulation to capture the clumpy ISM suggests that scattering off of GMCs after birth contributes to the heating, and this hypothesis is consistent with the measured  heating rate.  Given that \SIM\ has a very quiescent merger history, mergers are not likely to play a role in this heating.} 

\item{One of the key criteria necessary to match the MW's observed AVR is the ability to form stars with a low velocity dispersion, \vbirth, in order to match the low velocity dispersions seen in young stars in the MW.  This simulation follows the non-equilibrium formation and destruction of molecular hydrogen H$_2$ gas, and ensures that stars form only in the presence of H$_2$  \citep{Christensen2012}. Importantly, the stars are then born with this low dispersion, and not prone to processes that can introduce additional sources of dispersion at birth, such as star formation in outflowing gas \citep{ElBadry2016}. The ability to model cold gas velocity dispersions is also what allows the upside-down disk formation, as well as subsequent heating, to emerge. }

\item{Not only does \SIM\ match the local observations of the Milky Way, but its evolution is also completely consistent with observations that show that disks were kinematically hotter in the past and have ``settled'' with time.  We have shown that a model that has been invoked to explain disk settling \citep{Wisnioski2015}, in which the disk evolves in quasi-equilibrium and the velocity dispersion of the gas is tied to the evolving gas fraction inside disk galaxies, can match the evolution of \vbirth\ seen in \SIM\ (Figure \ref{fig:highz}).  Thus, the equilibrium model can explain both the populations of disk galaxies with time, and the evolution of this individual galaxy.}

\end{itemize}
\SIM\ is just one simulation of a Milky Way-mass galaxy, so it is not yet clear if the results will persist in additional simulations.  However, \citet{Kassin2014} showed that four simulated galaxies, including \SIM, matched the declining gas velocity dispersion witnessed over time in galaxy disks, suggesting that other galaxies may show similar trends.  In future work, we will explore a larger range of merger histories to discover if the match to the MW AVR can hold in a less quiescent galaxy. 

The correlation between \vbirth\ and \sfrden\ across both spatial position and time noted in this work (Figure~\ref{fig:radprofs}) is expected for galaxies on the star-forming main sequence and generically predicts upside-down disk formation.
Most theoretical models of star formation -- marginal gas disk stability including self-gravity considerations \citep[\eg][]{Toomre1964}, feedback driven turbulence \citep[\eg][]{Faucher-Giguere2013}, and accretion powered turbulence \citep[\eg][]{Krumholz2018} -- ultimately predict a correlation between gas velocity dispersion and \sfrden.
If these models capture reality, the velocity dispersion of the star-forming gas reservoir will decrease with time in a galaxy with a declining SFH, just as in \SIM. If \sfrden\ has a negative radial gradient, then \vbirth\ will decrease with time even for an unchanging SFR. As the galaxy grows inside-out, the characteristic radius of the star-forming disk moves outwards to lower \sfrden\ and thus lower \vbirth.   

Additionally, we have shown in this work that a quasi-stable model for \SIM's disk can predict the evolution in \vbirth\ seen in the simulation, but we have not fully verified that the model describes the simulation's underlying physics.  In a future work, we will test the equilibrium of \SIM's disk.  We will also explore the heating mechanism of stars after birth, and use the kinematics of simulated stars (e.g., the shape of the velocity ellipsoid) to interpret whether similar mechanisms might have been at play in shaping the Milky Way's AVR.

%

\bigskip
\section*{Acknowledgements}
We thank Stelios Kazantzidis for valuable discussions and insights during earlier phases of this study.
Support for SL was provided by NASA through Hubble Fellowship grant \#HST-JF2-51395.001-A awarded by the Space Telescope Science Institute, which is operated by the Association of Universities for Research in Astronomy, Inc., for NASA, under contract NAS5-26555.
This research was supported in part by the National Science Foundation under Grant No. NSF PHY-1748958 (KITP). Resources supporting this work were provided by the NASA High-End Computing (HEC) Program through the NASA Advanced Supercomputing (NAS) Division at Ames Research Center. 


\software {astropy \citep{astropy}}






\bibliographystyle{aasjournal.bst}
\bibliography{refs}



\end{document}